\documentclass[sigconf]{acmart}
\usepackage{amsmath}
\usepackage{color}
\usepackage{array}
\usepackage{{inputenc}}
\usepackage{balance}
\usepackage{multirow,tabularx}
\usepackage{booktabs,longtable}
\usepackage{algorithm}
\usepackage{algorithmic}
\usepackage{graphicx}
\usepackage{enumitem}
\usepackage{verbatim}
\usepackage{amsfonts}
\usepackage{hyperref}
\hypersetup{colorlinks=false,linkcolor=blue,urlcolor=blue,citecolor=red}
\usepackage{subcaption}

\newcommand\norm[1]{\left\lVert#1\right\rVert}
\newcommand{\Lapl}{\mathbf{\mathop{\mathcal{L}}}}

\newcommand{\Trans}[1]{{#1}^{\top}}

\newcommand{\Mat}[1]{\textbf{#1}}

\newcommand{\Space}[1]{\mathbb{#1}}
\newcommand{\Set}[1]{\mathcal{#1}}

\newcommand{\ie}{\emph{i.e., }}
\newcommand{\eg}{\emph{e.g., }}

\newcommand{\wrt}{\emph{w.r.t. }}
\newcommand{\cf}{\emph{cf. }}

\AtBeginDocument{%
  \providecommand\BibTeX{{%
    \normalfont B\kern-0.5em{\scshape i\kern-0.25em b}\kern-0.8em\TeX}}}

\copyrightyear{2021}
\acmYear{2021}
\setcopyright{iw3c2w3}
\acmConference[WWW '21]{Proceedings of the Web Conference 2021}{April 19--23, 2021}{Ljubljana, Slovenia}
\acmBooktitle{Proceedings of the Web Conference 2021 (WWW '21), April 19--23, 2021, Ljubljana, Slovenia}
\acmPrice{}
\acmDOI{10.1145/3442381.3450133}
\acmISBN{978-1-4503-8312-7/21/04}

\begin{document}

% \settopmatter{printacmref=false}

\title{Learning Intents behind Interactions with Knowledge Graph\\   for Recommendation}
% \title{On Relational Modeling in Knowledge-aware Recommendation}
% \title{Interpreting User Intents in Knowledge-aware Recommendation}
% \title{Interpreting User Intents in Knowledge-aware Recommendation}
% \title{On Relational Modeling for Knowledge-aware Recommendation}
% \title{Relational Modeling over Collaborative and Knowledge Graphs for Recommendation}

\author{Xiang Wang$^{1}$, Tinglin Huang$^{2}$, Dingxian Wang$^{3}$, Yancheng Yuan$^{4}$, Zhenguang Liu$^{2}$,\\Xiangnan He$^{5*}$\authornote{Xiangnan He is the corresponding author. The first three authors contribute equally.}, Tat-Seng Chua$^{1}$}
\affiliation{%
	\institution{\textsuperscript{1}National University of Singapore, \textsuperscript{2}Zhejiang University, \textsuperscript{3}eBay, \\\textsuperscript{4}The Hong Kong Polytechnic University, \textsuperscript{5}University of Science and Technology of China}
}
\email{xiangwang@u.nus.edu, tinglin.huang@zju.edu.cn, diwang@ebay.com}
\email{{yanchengyuanmath,liuzhenguang2008,xiangnanhe}@gmail.com, dcscts@nus.edu.sg}
\renewcommand{\authors}{Xiang Wang, Tinglin Huang, Dingxian Wang, Yancheng Yuan, Zhenguang Liu, Xiangnan He, Tat-Seng Chua}
\renewcommand{\shortauthors}{Wang~\textit{et al.}}

% \author{Xiang Wang$^{1}$, Tinglin Huang$^{2}$, Dingxian Wang$^{3}$, Yancheng Yuan$^{4}$, Zhenguang Liu$^{2}$,\\Xiangnan He$^{1}$\authornotemark[*], Tat-Seng Chua$^{5}$}
% \affiliation{%
% 	\institution{\textsuperscript{1}University of Science and Technology of China \textsuperscript{2}Zhejiang University, \textsuperscript{3}eBay, \\\textsuperscript{4}The Hong Kong Polytechnic University, \textsuperscript{5}National University of Singapore,}
% }
% \email{xiangwang@u.nus.edu, tinglin.huang@zju.edu.cn, diwang@ebay.com}
% \email{{yanchengyuanmath,liuzhenguang2008,xiangnanhe}@gmail.com, dcscts@nus.edu.sg}
% \authornote{Xiangnan He is the corresponding author.}
% \renewcommand{\authors}{Xiang Wang, Tinglin Huang, Dingxian Wang, Yancheng Yuan, Zhenguang Liu, Xiangnan He, Tat-Seng Chua}

%%
%% The abstract is a short summary of the work to be presented in the
%% article.
\begin{abstract}
  Knowledge graph (KG) plays an increasingly important role in recommender systems.
  A recent technical trend is to develop end-to-end models founded on graph neural networks (GNNs).
  However, existing GNN-based models are coarse-grained in relational modeling, failing to (1) identify user-item relation at a fine-grained level of intents, and (2) exploit relation dependencies to preserve the semantics of long-range connectivity.

  In this study, we explore intents behind a user-item interaction by using auxiliary item knowledge, and propose a new model, \emph{Knowledge Graph-based Intent Network} (KGIN).
  Technically, we model each intent as an attentive combination of KG relations, encouraging the independence of different intents for better model capability and interpretability.
  Furthermore, we devise a new information aggregation scheme for GNN, which recursively integrates the relation sequences of long-range connectivity (\ie relational paths).
  This scheme allows us to distill useful information about user intents and encode them into the representations of users and items.
  Experimental results on three benchmark datasets show that, KGIN achieves significant improvements over the state-of-the-art methods like KGAT~\cite{KGAT}, KGNN-LS~\cite{KGNN-LS}, and CKAN~\cite{CKAN}. Further analyses show that KGIN offers interpretable explanations for predictions by identifying influential intents and relational paths. The implementations are available at \url{https://github.com/huangtinglin/Knowledge_Graph_based_Intent_Network}.

\end{abstract}




\maketitle

\section{Introduction}

%% TODO: Background of knowledge-aware recommendation. At the core is to exploit knowledge graph to interpret/model user-item relationships.
Knowledge graph (KG) has shown great potential in improving the accuracy and explainability of recommendation.
The rich entity and relation information in KG can supplement the relational modeling between users and items. 
% At the core is to exploit its rich relational information to enhance the relationships between users and items, and further interpret user preference on items.
% Specifically, KG offers auxiliary facts about items in a collection of triplets, where each triplet $(h,r,t)$ describes a relation $r$ from a head entity $h$ to a tail entity $t$.
% heterogeneous graph, where each edge describes a relation between two real-world entities.
They not only reveal various relatedness among items (\eg co-directed by a person), but also can be used to interpret user preference (\eg attributing a user's choice of a movie to its director).

\begin{figure}[t]
    \centering
	\includegraphics[width=0.47\textwidth]{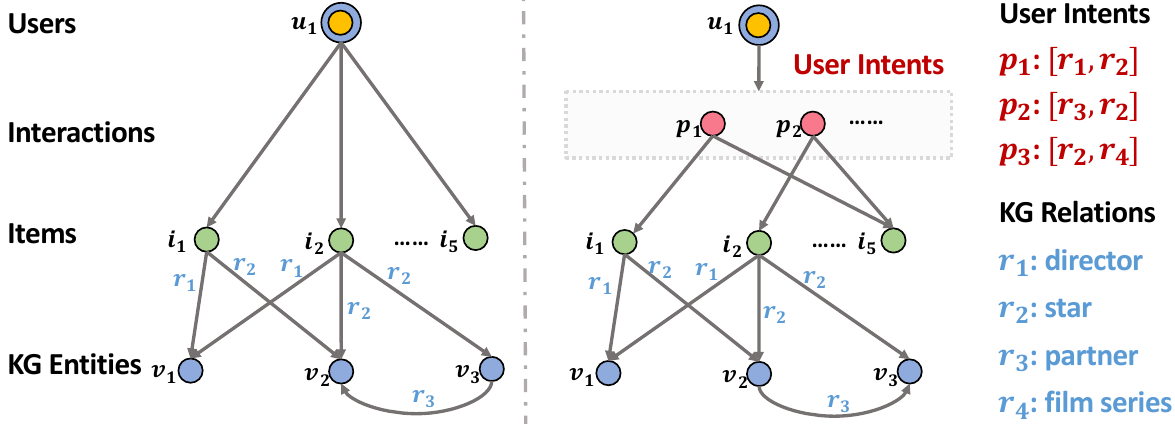}
	\vspace{-10pt}
	\caption{An example of user intents on adopting items (\ie fine-grained preference), where an arrow is the relation from a head entity to a tail entity. Best viewed in color.}
	\label{fig:intro-intent}
	\vspace{-10pt}
\end{figure}

%% TODO: Towards modeling of user-item relationships, related works have been proposed.
Learning high-quality user and item representations from such structural knowledge has become the theme of knowledge-aware recommendation.
% Modeling such structural knowledge is becoming the theme of knowledge-aware recommendation, especially to enhance the learning of user and item representations.
Earlier works~\cite{CKE,CFKG,KTUP} generate embeddings from KG triplets and treat them as prior or content information to supplement item representations.
Some follow-on studies~\cite{KPRN,hu2018leveraging,PGPR} enrich the interactions with multi-hop paths from users to items for better characterizing user-item relations.
However, these methods struggle to obtain high-quality paths, suffering from various issues like labor-intensive feature engineering~\cite{KPRN}, poor transferability to different domains~\cite{hu2018leveraging,NIRec}, and/or unstable performance~\cite{PGPR}.
% However, these methods struggle to obtain high-quality paths, which severely limits the benefits of structural knowledge.
% Because the path extraction easily suffers from labor-intensive feature engineering~\cite{KPRN}, poor transferability to different domains~\cite{hu2018leveraging,NIRec}, or unstable performance~\cite{PGPR}.
More recently, a technical trend~\cite{KGAT,KGNN-LS,KGCN,CKAN} is to develop end-to-end models founded on graph neural networks (GNNs)~\cite{DBLP:conf/nips/HamiltonYL17,DBLP:conf/iclr/KipfW17,DBLP:conf/iclr/VelickovicCCRLB18,DBLP:journals/corr/abs-2003-00982}.
The key idea is to utilize the information aggregation scheme, which can effectively integrate multi-hop neighbors into representations.
Benefiting from the integration of connectivity modeling and representation learning, these GNN-based models achieve promising performance for recommendation.
% More recently, propagation-based works~\cite{KGAT,KGNN-LS,KGCN,CKAN} are attracting considerable attention and has achieved remarkable improvements.
% The basic idea is the information propagation mechanism of graph neural networks (GNNs)~\cite{DBLP:conf/nips/HamiltonYL17,DBLP:conf/iclr/KipfW17,DBLP:conf/iclr/VelickovicCCRLB18,DBLP:journals/corr/abs-2003-00982}, which incorporates information from multi-hop neighbors to encode the long-range connectivity into the representations.
% We benefit such success to the integration of connectivity modeling and representation learning.

\begin{figure}[t]
    \centering
	\includegraphics[width=0.48\textwidth]{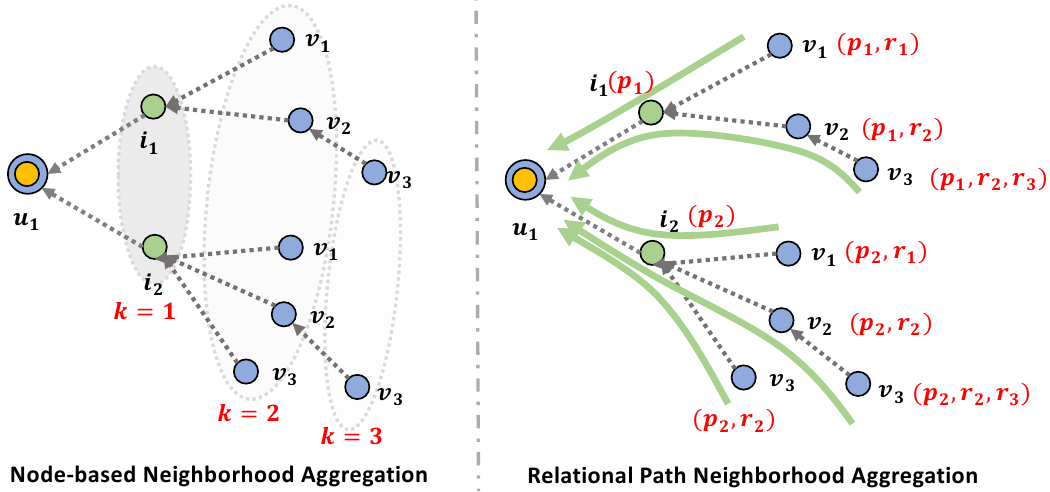}
	\vspace{-10pt}
	\caption{An example of the node-based and relational path-aware aggregation schemes, where a (dashed or solid) arrow is an information flow among nodes. Best viewed in color.}
	\label{fig:intro-aggregation}
	\vspace{-10pt}
\end{figure}

Despite their effectiveness, we argue that current GNN-based methods fall short in modeling two factors:
(1) \textbf{User Intents}.
To the best of our knowledge, none of these studies consider user-item relations at a finer-grained level of intents.
% intents (\ie fine-grained relationships) of users on adopting items.
% Only one coarse-grained relation is assumed to exist between users and items, which is the interaction behavior.
An important fact has been ignored: a user typically has multiple intents, driving the user to consume different items.
% They ignore an underlying fact that, the interaction behaviors are usually driven by different intents of users.
Taking the right of Figure~\ref{fig:intro-intent} as an example,
intent $p_1$ emphasizes a combination of \emph{director} ($r_{1}$) and \emph{star} ($r_{2}$) aspects that drives user $u_1$ to watch the movies $i_1$ and $i_5$; while another intent $p_2$ highlights the \emph{star} ($r_{2}$) and \emph{partner} $(r_{3})$ aspects the user to select movie $i_2$.
Ignoring the existence of user intents limits the modeling of user-item interactions.
(2) \textbf{Relational Paths}.
In these studies, the information aggregation schemes are mostly node-based --- that is, to collect the information from neighboring nodes,
without differentiating which paths it comes from.
Moreover, KG relations are typically modeled in decay factors~\cite{KGAT,KGNN-LS} of adjacent matrix, in order to control the influences of neighbors.
As shown in the left of Figure~\ref{fig:intro-aggregation}, the representation of $u_{1}$ mixes the signal from the one-, two-, and three-hop neighbors (\ie $\{i_1,i_2\}$, $\{v_1,v_2,v_3\}$, $\{v_{3}\}$, respectively).
It fails to preserve the relation dependencies and sequencies carried by paths (\eg $(p_{1},r_{2},r_{3})$ in the three-hop path from $v_{3}$ to $u_{1}$).
Hence, such node-based schemes are insufficient to capture the interactions among relations.
% and distill relational information pertinent to user intents.

% The propagation mechanisms in prior studies emphasize the information carried by neighboring nodes, but fail to preserve the semantics of relational paths.
% For example, item $i_1$'s representation aggregates the information of one-hop neighbors $\{e_1,e_2\}$ and two-hop neighbors $\{e_3,e_4\}$, without differentiating which paths they belong to.
% Hence, they fail to capture the holistic semantics of relational paths (sya, the combination of \emph{director} and \emph{genre} in path $p_1$).

In this paper, we focus on exploring user intents behind user-item interactions by using item KG, so as to improve the performance and interpretability of recommendation.
We propose a new model, \emph{Knowledge Graph-based Intent Network} (KGIN), which consists of two components to solve the foregoing limitations correspondingly:
(1) \textbf{User Intent Modeling}. Each user-item interaction is enriched with the underlying intents. While we can express these intents as latent vectors, their semantics are opaque to understand. Hence, we associate each intent with a distribution over KG relations, accounting for the importance of relation combination.
Technically, the intent embedding is an attentive combination of relation embeddings, where important relations are assigned with larger attribution scores.
Moreover, an independence constraint is introduced to  encourage significant differences among intents for better interpretability.
(2) \textbf{Relational Path-aware Aggregation}.
Unlike the node-based aggregation mechanism,
% Distinct from prior studies~\cite{KGAT,KGNN-LS} that simply use KG relations in the decay factors to control how much information are propagated from nodes, 
we view a relational path as an information channel, and embed each channel into a representation vector.
As user-intent-item triplets and KG triplets are heterogeneous, we set different aggregation strategies for these two parts, so as to distill behavioral patterns of users and relatedness of items respectively.
% As the user-intent-item triplets and KG triplets contain collaborative signals of users and relatedness signals of items respectively, we set heterogeneous propagations over these two parts.
In a nutshell, this relational modeling allows us to identify influential intents, and encode relation dependencies and semantics of paths into the representations.
We conduct extensive experiments on three real-world datasets. 
Experimental results show that our KGIN outperforms the state-of-the-art methods such as KGAT~\cite{KGAT}, KGNN-LS~\cite{KGNN-LS}, and CKAN~\cite{CKAN}.
Furthermore, KGIN is able to interpret user behaviors at the granularity of intents.

We summarize the contributions of this work as: 
% Contribution of this work can be summarized as:
\begin{itemize}[leftmargin=*]
    \item Revealing user intents behind the interactions in KG-based recommendation, for better model capacity and interpretability;
    % \item We focus on the relational modeling in the knowledge-aware recommendation, especially user intents and relational paths.
    \item Proposing a new model, KGIN, which considers user-item relationships at the finer granularity of intents and long-range semantics of relational paths under the GNN paradigm;
    \item Conducting empirical studies on three benchmark datasets to demonstrate the superiority of KGIN. 
    % \item We conduct extensive experiments on three benchmark datasets, demonstrating the effectiveness of KGIN and its explainability in understanding user intents.
\end{itemize}

\section{Problem Formulation}
We first introduce the structural data: user-item interactions and knowledge graph, and then formulate our task.

\vspace{5pt}
\noindent\textbf{Interaction Data.}
Here we focus on the implicit feedback~\cite{BPR} in recommendation, where the signal that a user provides about her preference is implicit (\eg view, click, purchase).
Let $\Set{U}$ be a set of users and $\Set{I}$ a set of items.
Let $\Set{O}^{+}=\{(u,i)|u\in\Set{U},i\in\Set{I}\}$ be a set of observed feedback, where each $(u,i)$ pair indicates that user $u$ has interacted with item $i$ before.
In some previous works like KGAT~\cite{KGAT}, an additional relation \emph{interact-with} is introduced to explicitly present the user-item relationship and convert a $(u,i)$ pair to the $(u,\emph{interact-with},i)$ triplet.
As such, the user-item interactions can be seamlessly combined with KG.

\vspace{5pt}
\noindent\textbf{Knowledge Graph.}
KG stores the structured information of real-world facts, such as item attributes, taxonomy, or external commonsense knowledge, in the form of a heterogeneous graph or heterogeneous information network~\cite{HIN,HIN-survey}.
Let $\Set{V}$ be a set of real-world entities and $\Set{R}$ be the relation set, which involves relations in both canonical and inverse directions (\eg \emph{director} and \emph{directed-by}).
Let $\Set{G}=\{(h,r,t)|h,t\in\Set{V},r\in\Set{R}\}$ be a collection of triplets, where each $(h,r,t)$ triplet indicates that a relation $r$ exists from head entity $h$ to tail entity $t$.
For example, (\emph{Martin Freeman, star, The Hobbit I}) describes that \emph{Martin Freeman} is the \emph{star} of movie \emph{The Hobbit I}.
With the mappings between items and KG entities ($\Set{I}\subset\Set{V}$), KG is able to profile items and offer complementary information to the interaction data.

\vspace{5pt}
\noindent\textbf{Task Description.}
Given the interaction data $\Set{O}^{+}$ and the KG $\Set{G}$, our task of knowledge-aware recommendation is to learn a function that can predict how likely a user would adopt an item.

% denotes the model for item recommendation. It takes a list of user-item pairs as input, and outputs a relevance score indicating the likelihood that u likes i, given the preference $p\in\Set{P}$, where the number of the preference set $\Set{P}$ is predefined.
% For each user-item pair, we induce a preference, serving as a similar role with the relation for two entities.

% \section{Preliminary and Related Work}
% \subsection{Structural Knowledge.}

% \noindent\textbf{Interaction Data}

% \noindent\textbf{Knowledge Graph}

% \noindent\textbf{Connectivity}

% \subsection{Knowledge-aware Recommendation.}

% \noindent\textbf{Embedding-based Methods}

% \noindent\textbf{Path-based Methods}

% \noindent\textbf{Policy-based Methods}

\section{Methodology}
\begin{figure*}[t]
    \centering
	\includegraphics[width=0.97\textwidth]{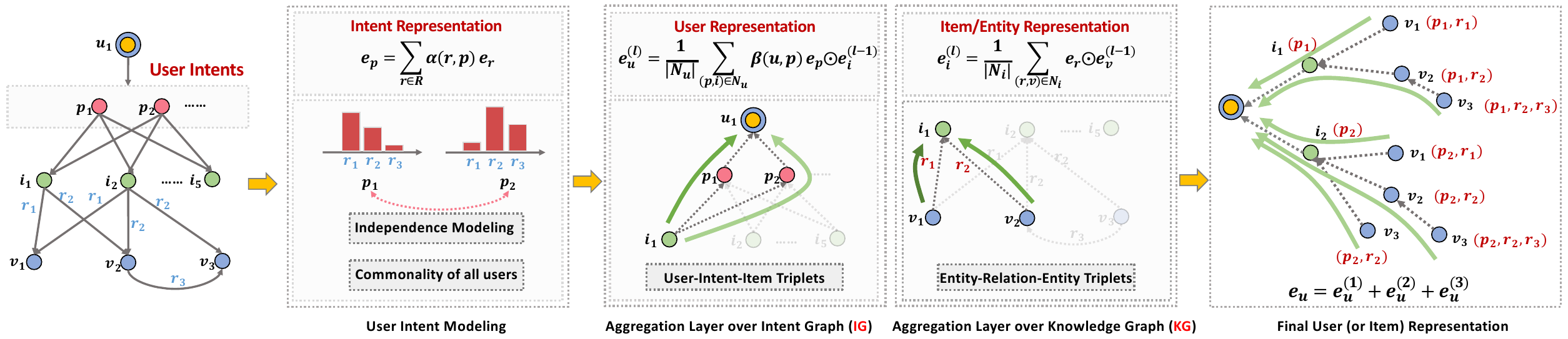}
	\vspace{-10pt}
	\caption{Illustration of the proposed KGIN framework. Best viewed in color.}
	\label{fig:framework}
	\vspace{-10pt}
\end{figure*}

We now present the proposed Knowledge Graph-based Intent Network (KGIN).
Figure~\ref{fig:framework} displays the working flow of KGIN.
It consists of two key components: (1) user intent modeling, which uses multiple latent intents to profile user-item relationships and formulates each intent as an attentive combination of KG relations, meanwhile encourages different intents to be independent of each others;
and (2) relational path-aware aggregation, which highlights the relation dependence in long-range connectivity, so as to preserve the holistic semantics of relational paths.
KGIN ultimately yields high-quality representations of users and items.

\subsection{User Intent Modeling}
Unlike the previous GNN-based studies~\cite{KGAT,KGNN-LS,CKAN} that assume no or only one \emph{interact-with} relation between users and items, we aim to capture the intuition that behaviors of users are influenced by multiple intents.
Here we frame the intent as the reason of users' choices to items, which reflects the commonality of all users' behaviors.
Taking movie recommendation as an example, possible intents are diverse considerations on movie attributes, such as the combination of \emph{star} and \emph{partner}, or \emph{director} and \emph{genre}.
Different intents abstract different behavioral patterns of users.
This can supercharge the widely-used collaborative filtering~\cite{BPR} effect with the finer-grained assumption --- users driven by similar intents would have similar preference on items.
Such intuition motivates us to model user-item relations at the granularity of intents.

% Distinct from previous studies~\cite{KGAT,KGNN-LS,CKAN} that assume only one \emph{interact-with} relation between users and items, behaviors of users are influenced by multiple intents.
% Taking movie recommendation as an example, user $u$ watches movie $i_1$ since the star matches her interest well, while caring more about the director and genre information when watching movie $i_2$. 
% Such intuition inspires us to model finer-grained relationships between users and items, especially at the granularity of intents.

Assuming $\Set{P}$ as the set of intents shared by all users, we can slice a uniform user-item relation into the $|\Set{P}|$ intents, and decompose each $(u,i)$ pair into $\{(u,p,i)|p\in\Set{P}\}$.
As a result, we reorganize the user-item interaction data as a heterogeneous graph, termed \textbf{intent graph (IG)}, which differs from the homogeneous collaborative graph adopted in previous works~\cite{KGAT,CKAN}.

% Intuitively, one user behavior is influenced by multiple intents, such as passing time, matching particular interests, and shopping for others like family. Taking movie recommendation as an example, user u passed time with movie i 1 , hence might care less about whether i 1 ’s director matches her interests well; whereas, u watched i 2 since its director is an important factor of u’s interest. Clearly, different intents have varying contributions to motivate user behaviors.

\subsubsection{\textbf{Representation Learning of Intents}}
Although we can express these intents with latent vectors, it is hard to identify the semantics of each intent explicitly.
One straightforward solution is to couple each intent with one KG relation, which is proposed by KTUP~\cite{KTUP}.
However, this solution only considers single relations in isolation, without accounting for the interactions and combinations of relations, thereby fails to refine high-level concepts of user intents.
For example, the combination of relations $r_1$ and $r_2$ is influential to intent $p_{1}$, while relations $r_3$ and $r_4$ contributes more to intent $p_{2}$.
Hence, we assign each intent $p\in\Set{P}$ with a distribution over KG relations --- technically, exert an attention strategy to create the intent embedding as:
\begin{gather}\label{equ:intent-embedding}
    \Mat{e}_{p}=\sum_{r\in\Set{R}}\alpha(r,p)\Mat{e}_{r},
\end{gather}
where $\Mat{e}_{r}$ is the ID embedding of relation $r$, which is assigned with an attention score $\alpha(r,p)$ to quantify its importance, formally:
\begin{gather}
    \alpha(r,p) = \frac{\exp(w_{rp})}{\sum_{r'\in\Set{R}}\exp(w_{r'p})},
\end{gather}
where $w_{rp}$ is a trainable weight specific to certain relation $r$ and certain intent $p$.
Here we use the weights for simplicity, and leave the further exploration of complex attention modules in future work.
It is worthwhile mentioning that the attentions are not tailored to a single user, but refine common patterns of all users.

\subsubsection{\textbf{Independence Modeling of Intents}}\label{sec:independence-of-intents}
Different intents should contain different information about user preference \cite{DisenGCN,MacridVAE}.
If one intent can be inferred by the others, it might be redundant and less informative to describe user-item relation; in contrast, the intent with unique information will offer a useful angle to characterize behavioral patterns of users.
Hence, for better model capacity and explainability, we encourage the representations of intents to differ from each others.

% Intuitively, different intents should contain varying information about user preference \cite{DisenGCN,MacridVAE}.
% Hence, the embedding of an intent is encouraged to integrate the unique and strongly relevant information about this kind of user-item relationship, and thus differ from the others'.
% For example, if one intent embedding $\Mat{e}_{p}$ can be inferred by the others $\{\Mat{e}_{p'}|k'\neq k\}$, the intent $p$ might be redundant and less informative to describe user-item relationships.

Here we introduce a module of independence modeling to guide the representation learning of independent intents.
This module can be simply achieved by applying a statistical measure, such as mutual information~\cite{DBLP:conf/icml/BelghaziBROBHC18}, Pearson correlation~\cite{szekely2007measuring}, and distance correlation~\cite{szekely2009brownian,szekely2007measuring,DGCF}, as a regularizer.
Here we offer two implementations:
\begin{itemize}[leftmargin=*]
    \item \textbf{Mutual information.} We minimize the mutual information between the representations of any two different intents, so as to quantify their independence. Such an idea coincides with contrastive learning \cite{DBLP:conf/icml/ChenK0H20,DBLP:journals/jmlr/GutmannH10}. More formally, the independence modeling is:
    \begin{gather}\label{equ:independence-loss-mi}
        \Lapl_{\text{IND}}=\sum_{p\in\Set{P}}-\log{\frac{\exp{(s(\Mat{e}_{p},\Mat{e}_{p})/\tau)}}{\sum_{p'\in\Set{P}}\exp{(s(\Mat{e}_{p},\Mat{e}_{p'})/\tau)}}},
    \end{gather}
    where $s(\cdot)$ is the function measuring the associations of any two intent representations, which is set as cosine similarity function here; and $\tau$ is the hyper-parameter to the temperature in softmax function.
    
    \item \textbf{Distance correlation.} It measures both linear and nonlinear associations of any two variables, whose coefficient is zero if and only if these variables are independent. Minimizing the distance correlation of user intents enables us to reduce the dependence of different intents, which is formulated as:
    \begin{gather}\label{equ:independence-loss-dc}
        \Lapl_{\text{IND}}=\sum_{p,p'\in\Set{P},~p\neq p'}dCor(\Mat{e}_{p},\Mat{e}_{p'}),
    \end{gather}
    where $dCor(\cdot)$ is the distance correlation between intents $p$ and $p'$:
    \begin{gather}
        dCor(\Mat{e}_{p},\Mat{e}_{p'})=\frac{dCov(\Mat{e}_{p},\Mat{e}_{p'})}{\sqrt{dVar(\Mat{e}_{p})\cdot dVar(\Mat{e}_{p'})}},
    \end{gather}
    where $dCov(\cdot)$ is the distance covariance of two representations, and $dVar(\cdot)$ is the distance variance of each intent representation.
\end{itemize}
Optimizing this loss allows us to encourage the divergence among different intents and makes these intents have distinct boundary, thus endows better explainability of user intents.

% \subsection{Relational Propagation Mechanism}
\subsection{Relational Path-aware Aggregation}
Having modeled the user intents, we move on to the representation learning of users and items under the GNN-base paradigm.
Previous GNN-based recommender models~\cite{KGAT,KGNN-LS,KGCN} have shown that the neighborhood aggregation scheme is a promising end-to-end way to integrate multi-hop neighbors into representations.
More specifically, the representation vector of an ego node is computed by recursively aggregating and transforming representations of its multi-hop neighbors.

However, we argue that current aggregation schemes are mostly node-based, which limit the benefit of the structural knowledge, due to two issues:
(1) The aggregators focus on combining the information of neighborhood, without distinguishing which paths they originate from.
Considering the example in Figure~\ref{fig:intro-aggregation}, there are three information channels between the ego node $u_{1}$ and its 2-hop neighbor $v_{2}$: $u_{1}\xleftarrow{p_{1}}i_{1}\xleftarrow{r_{2}}v_{2}$ and $u_{1}\xleftarrow{p_{2}}i_{2}\xleftarrow{r_{2}}v_{2}$.
When constructing the neural messages passed by $v_{2}$, the node-based aggregators largely transform and rescale $v_{2}$'s representation by decay factors, without considering influences of different channels.
Hence, they are insufficient to preserve the structural information in representations.
Moreover, (2) current node-based aggregators usually model KG relations in the decay factors via attention networks~\cite{CKAN,KGAT,KGNN-LS} to control how much information is propagated from neighbors.
This limits the contributions of KG relations to node representations.
Moreover, no relation dependency (\eg $(p_{2},r_{2},r_{3})$ in path $u_{1}\xleftarrow{p_{2}}i_{2}\xleftarrow{r_{2}}v_{2}\xleftarrow{r_{3}}v_{3}$) is captured in an explicit fashion.
Hence, we aim to devise a relational path-aware aggregation scheme to solve these two limitations.

\subsubsection{\textbf{Aggregation Layer over Intent Graph}}
We first move on to refine collaborative information from IG.
As mentioned, the CF effect~\cite{BPR} is powerful to characterize user patterns, by assuming that behavioral similar users would have similar preference on items.
This inspires us to treat personal history (\ie the items a user has adopted before) as the pre-existing features of individual users.
Moreover, in our IG, we can capture finer-grained patterns at a granular level of user intents, by assuming that users with similar intents would exhibit similar preference towards items.
Considering a user $u$ in IG, we use $\Set{N}_{u}=\{(p,i)|(u,p,i)\in\Set{C}\}$ to represent the intent-aware history and the first-order connectivity around $u$.
Technically, we can integrate the intent-aware information from historical items to create the representation of user $u$ as:
\begin{gather}\label{equ:ig-aggregation}
    \Mat{e}^{(1)}_{u}=f_{\text{IG}}\Big(\big\{(\Mat{e}^{(0)}_{u},\Mat{e}_{p},\Mat{e}^{(0)}_{i})|(p,i)\in\Set{N}_{u}\big\}\Big),
\end{gather}
where $\Mat{e}^{(1)}_{u}\in\Space{R}^{d}$ is the representation of user $u$; and $f_{\text{IG}}(\cdot)$ is the aggregator function to characterize each first-order connection $(u,p,i)$.
Here we implement $f_{\text{IG}}(\cdot)$ as:
\begin{gather}\label{equ:ig-aggregation-implement}
    \Mat{e}^{(1)}_{u} = \frac{1}{|\Set{N}_{u}|}\sum_{(p,i)\in\Set{N}_{u}}\beta(u,p)\Mat{e}_{p}\odot\Mat{e}^{(0)}_{i},
\end{gather}
where $\Mat{e}^{(0)}_{i}$ is the ID embedding of item $i$; $\odot$ is the element-wise product.
We harness it with two insights.
(1) For a given a user, different intents will have varying contributions to motivate her behaviors.
Hence, we introduce an attention score $\beta(u,p)$ to differentiate the importance of intent $p$ as:
\begin{gather}\label{equ:intent-attention}
    \beta(u,p) = \frac{\exp(\Trans{\Mat{e}}_{p}\Mat{e}^{(0)}_{u})}{\sum_{p'\in\Set{P}}\exp(\Trans{\Mat{e}}_{p'}\Mat{e}^{(0)}_{u})},
\end{gather}
where $\Mat{e}^{(0)}_{u}\in\Space{R}^{d}$ is the ID embedding of user $u$ to make the importance score personalized.
(2) Unlike the ideas of using the decay factors~\cite{KGAT,CKAN,KGNN-LS} or regularization terms~\cite{KGAT} in previous studies, we highlight the role of intent relations during the aggregation. Hence we construct the item $i$'s message via the element-wise product $\beta(u,p)\Mat{e}_{p}\odot\Mat{e}^{(0)}_{i}$.
As a result, we are able to explicitly express the first-order intent-aware information in the user representations.

\subsubsection{\textbf{Aggregation Layer over Knowledge Graph}}
We then focus on the aggregation scheme in KG.
As one entity can be involved in multiple KG triplets, it can take other connected entities as its attributes, which reflect the content similarity among items.
For example, movie \emph{The Hobbit I} can be described by its \emph{director} \emph{Peter Jackson} and \emph{star} \emph{Martin Freeman}.
More formally, we use $\Set{N}_{i}=\{(r,v)|(i,r,v)\in\Set{G}\}$ to represent the attributes and the first-order connectivity about item $i$, and then integrate the relation-aware information from connected entities to generate the representation of item $i$:
\begin{gather}\label{equ:kg-aggregation}
    \Mat{e}^{(1)}_{i}=f_{\text{KG}}\Big(\{(\Mat{e}^{(0)}_{i},\Mat{e}_{r},\Mat{e}^{(0)}_{v})|(r,v)\in\Set{N}_{i}\}\Big)
\end{gather}
where $\Mat{e}^{(1)}_{i}\in\Space{R}^{d}$ is the representation collecting the information from the first-order connectivity; and $f_{\text{KG}}(\cdot)$ is the aggregation function to extract and integrate information from each connection $(i,r,v)$.
Here we account for the relational context in the aggregator.
Intuitively, each KG entity has different semantics and meanings in different relational contexts.
For instance, entity \emph{Quentin Tarantino} expresses signals pertinent to the \emph{director} and \emph{star} concepts in two triplets (\emph{Quentin Tarantino}, \emph{director}, \emph{Django Unchained}) and (\emph{Quentin Tarantino}, \emph{star}, \emph{Django Unchained}), respectively.
However, previous studies~\cite{KGAT,CKAN,KGNN-LS} only model KG relations in the decay factors via the attention mechanism, in order to control the contributions of \emph{Quentin Tarantino} to the representation of \emph{Django Unchained}.
Instead, we model the relational context in the aggregator as:
\begin{gather}\label{equ:kg-aggregation-implement}
    \Mat{e}^{(1)}_{i}=\frac{1}{|\Set{N}_{i}|}\sum_{(r,v)\in\Set{N}_{i}}\Mat{e}_{r}\odot\Mat{e}^{(0)}_{v},
\end{gather}
where $\Mat{e}^{(0)}_{v}$ is the ID embedding of entity $v$.
For each triplet $(i,r,v)$, we devise a relational message $\Mat{e}_{r}\odot\Mat{e}^{(0)}_{v}$ by modeling the relation $r$ as the projection or rotation operator~\cite{RotatE}.
As a result, the relational message is able to reveal different meanings carried by the triplets, even when they get the same entities.
Analogously, we can obtain the representation $\Mat{e}^{(1)}_{v}$ of each KG entity $v\in\Set{V}$.

\subsubsection{\textbf{Capturing Relational Paths}}
Having modeled the first-order connectivity in Equations \eqref{equ:ig-aggregation} and \eqref{equ:kg-aggregation}, we further stack more aggregation layers to gather the influential signals from higher-order neighbors.
Technically, we recursively formulate the representations of user $u$ and item $i$ after $l$ layers as:
\begin{align}\label{equ:multiple-layer}
    \Mat{e}^{(l)}_{u}&=f_{\text{IG}}\Big(\big\{(\Mat{e}^{(l-1)}_{u},\Mat{e}_{p},\Mat{e}^{(l-1)}_{i})|(p,i)\in\Set{N}_{u}\big\}\Big),\nonumber\\
    \Mat{e}^{(l)}_{i}&=f_{\text{KG}}\Big(\{(\Mat{e}^{(l-1)}_{i},\Mat{e}_{r},\Mat{e}^{(l-1)}_{v})|(r,v)\in\Set{N}_{i}\}\Big),
\end{align}
where $\Mat{e}^{(l-1)}_{u}$, $\Mat{e}^{(l-1)}_{i}$, $\Mat{e}^{(l-1)}_{v}$ separately denote the representations of user $u$, item $i$, and entity $v$, which memorize the relational signals being propagated from their ($l$-1)-hop neighbors.
Benefiting from our relational modeling, these representations are able to store the holistic semantics of multi-hop paths, and highlight the relational dependencies.
Let $s=i\xrightarrow{r_{1}}s_{1}\xrightarrow{r_{2}}\cdots s_{l-1}\xrightarrow{r_{l}}s_{l}$ be a $l$-hop path rooted at item $i$, which contains a sequence of connected triplets.
Its relational path is represented as the sequence of relations merely, \ie $(r_{1},r_{2},\cdots,r_{l})$.
We can rewrite the representation $\Mat{e}^{(l)}_{i}$ as follows:
\begin{gather}\label{equ:relational-path-modeling}
    \Mat{e}^{(l)}_{i} = \sum_{s\in\Set{N}_{i}^{l}}\frac{\Mat{e}_{r_{1}}}{|\Set{N}_{s_{1}}|}\odot\frac{\Mat{e}_{r_{2}}}{|\Set{N}_{s_{2}}|}\odot\cdots\odot\frac{\Mat{e}_{r_{l}}}{|\Set{N}_{s_{l}}|}\odot\Mat{e}^{(0)}_{s_{l}},
\end{gather}
where $\Set{N}^{l}_{i}$ is the set of all $i$'s $l$-hop paths.
Clearly, this representation reflects the interactions among relations and preserves the holistic semantics of paths.
This is significantly different from the current aggregation mechanism adopted in knowledge-aware recommenders, which overlook the importance of KG relations and thus fail to capture the relational paths.

% \subsubsection{\textbf{Layer Combination}}

\subsection{Model Prediction}
After $L$ layers, we obtain the representations of user $u$ and item $i$ at different layers and then sum them up as the final representations:
\begin{gather}
    \Mat{e}^{*}_{u}=\Mat{e}^{(0)}_{u}+\cdots+\Mat{e}^{(L)}_{u},\quad\quad\Mat{e}^{*}_{i}=\Mat{e}^{(0)}_{i}+\cdots+\Mat{e}^{(L)}_{i}.
\end{gather}
By doing so, the intent-aware relationships and the KG relation dependencies of paths are encoded in the final representations.

Thereafter, we employ the inner product on the user and item representations to predict how likely the user would adopt the item:
\begin{gather}
    \hat{y}_{ui}=\Trans{\Mat{e}^{*}_{u}}\Mat{e}^{*}_{i}.
\end{gather}

\subsection{Model Optimization}

We opt for the pairwise BPR loss~\cite{BPR} to reconstruct the historical data.
Specifically, it considers that for a given user, her historical items should be assigned with higher prediction scores than the unobserved items:
\begin{gather}
    \Lapl_{\text{BPR}}=\sum_{(u,i,j)\in\Set{O}}-\ln\sigma(\hat{y}_{ui}-\hat{y}_{uj}),
\end{gather}
where $\Set{O}=\{(u,i,j)|(u,i)\in\Set{O}^{+},(u,j)\in\Set{O}^{-}\}$ is the training dataset consisting of the observed interactions $\Set{O}^{+}$ and unobserved counterparts $\Set{O}^{-}$; $\sigma(\cdot)$ is the sigmoid function.
By combining the independence loss and BPR loss, we minimize the following objective function to learn the model parameter:
\begin{gather}
    \Lapl_{\text{KGIN}} = \Lapl_{\text{BPR}} + \lambda_{1}\Lapl_{\text{IND}} + \lambda_{2}\norm{\Theta}^{2}_{2},
\end{gather}
where $\Theta=\{\Mat{e}^{(0)}_{u},\Mat{e}^{(0)}_{v},\Mat{e}_{r},\Mat{e}_{p}, \Mat{w}|u\in\Set{U},v\in\Set{V},p\in\Set{P}\}$ is the set of model parameters (note that the item set $\Set{I}\subset\Set{V}$); $\lambda_{1}$ and $\lambda_{2}$ are two hyperparameters to control the independence loss (Equation~\eqref{equ:ig-aggregation}) and $L_{2}$ regularization term, respectively.

\subsection{Model Analysis}
% We analyze our KGIN model from multiple dimensions: model size, time complexity, and relations with other models.

\subsubsection{\textbf{Model Size}}
Recent studies~\cite{SGCN} have shown that using nonlinear feature transformations possibly makes GNNs difficult to train.
Hence, in the aggregation scheme of KGIN, we discard the nonlinear activation functions and feature transformation matrices.
Hence, the model parameters of KGIN consist of (1) ID embeddings of users, KG entities (including items), and KG relations $\{\Mat{e}^{(0)}_{u},\Mat{e}^{(0)}_{v},\Mat{e}_{r}|u\in\Set{U},v\in\Set{V}\}$; and (2) ID embeddings of user intents $\{\Mat{e}_{p}|p\in\Set{P}\}$ and the attention weights $\Mat{w}$.

\subsubsection{\textbf{Time Complexity}}
The time cost of KGIN mainly comes from the user intent modeling and aggregation scheme.
In the aggregations over IG, the computational complexity of user representations is $O(L|\Set{C}|d)$, where $L$, $|\Set{C}|$, and $d$ denote the number of layers, the number of triplets in IG, and the embedding size, respectively.
In the aggregation over KG, the time cost of updating entity representations is $O(L|\Set{G}|d)$, where $|\Set{G}|$ is the number of KG triplets.
As for the independence modeling, the cost of distance correlation is $O(|\Set{P}|(|\Set{P}|-1)/2)$, where $|\Set{P}|$ is the number of user intents.
In total, the time complexity of the whole training epoch is $O(L|\Set{C}|d+L|\Set{G}|d+|\Set{P}|(|\Set{P}|-1)/2)$.
Under the same experimental settings (\ie representation sizes at different layers), KGIN has comparable complexity to KGAT and CKAN.

\section{Experiments}
We provide empirical results to demonstrate the effectiveness of our proposed KGIN. The experiments are designed to answer the following research questions:
\begin{itemize}[leftmargin=*]
    \item \textbf{RQ1:} How does KGIN perform, comparing to the state-of-the-art knowledge-aware recommender models?
    \item \textbf{RQ2:} What is the impact of the designs (\eg the number and independence of user intents, the depth of relational paths) on the improvement of KGIN's relational modeling?
    \item \textbf{RQ3:} Can KGIN provide insights on user intents and give an intuitive impression of explainability?
\end{itemize}

\subsection{Experimental Settings}
\subsubsection{\textbf{Dataset Description}}

\begin{table}[t]
    \caption{Statistics of the datasets.}
    \vspace{-10px}
    \label{tab:dataset}
    \resizebox{0.48\textwidth}{!}{
    \begin{tabular}{c|l|r|r|r}
    \hline
    \multicolumn{1}{l|}{} &  & \multicolumn{1}{c|}{Amazon-Book} & \multicolumn{1}{c|}{Last-FM} & \multicolumn{1}{c}{Alibaba-iFashion} \\ \hline\hline
    \multirow{3}{*}{\begin{tabular}[c]{@{}c@{}}User-Item\\ Interaction\end{tabular}} & \#Users & 70,679 & 23,566 & 114,737 \\
     & \#Items & 24,915 & 48,123 & 30,040 \\ 
     & \#Interactions & 847,733 & 3,034,796 & 1,781,093 \\ \hline\hline
    \multirow{3}{*}{\begin{tabular}[c]{@{}c@{}}Knowledge\\ Graph\end{tabular}} & \#Entities & 88,572 & 58,266 & 59,156 \\
     & \#Relations & 39 & 9 & 51 \\
     & \#Triplets & 2,557,746 & 464,567 & 279,155 \\ \hline
    \end{tabular}}
    \vspace{-10px}
\end{table}

We use three benchmark datasets for book, music, and fashion outfit recommendation in the experiments:
(1) We use the Amazon-Book and Last-FM datasets released by KGAT~\cite{KGAT};
And (2) we further introduce the Alibaba-iFashion dataset~\cite{POG} to investigate the effectiveness of item knowledge.
This is a fashion outfit dataset collected from Alibaba online shopping systems. The outfits are viewed as the items being recommended to users, where each outfit consists of multiple fashion staffs (\eg tops, bottoms, shoes, accessories), and these staffs follow a fashion taxonomy and are assigned with different fashion categories (\eg jeans, T-shirts).
We extract such attributes as the KG data of outfits.
Moreover, in order to ensure the data quality, we adopt the 10-core setting, \ie discarding users and items with less than ten interactions, and filtering out KG entities involved less than ten triplets.
The statistics of datasets are summarized in Table~\ref{tab:dataset}, where we only list the number of canonical relations and construct triplets with the inverse relations in experiments.
Closely Following prior studies~\cite{KGAT,KGPolicy}, we use the same data partition.
In the training phase, each observed user-item interaction is a positive instance, while an item that the user did not adopt before is randomly sampled to pair the user as a negative instance.

\subsubsection{\textbf{Evaluation Metrics}}
In the evaluation phase, we conduct the all-ranking strategy~\cite{DBLP:conf/kdd/KricheneR20}, rather than sampled metrics like leaving one item out~\cite{KPRN} or sampling a smaller set of users~\cite{KGNN-LS,CKAN,KGCN}.
To be more specific, for each user, the full items that she has not adopted before are viewed as negative, and the relevant items in the testing set are treated as positive.
All these items are ranked based on the predictions of recommender model.
To evaluate top-$K$ recommendation, we adopt the protocols~\cite{DBLP:conf/kdd/KricheneR20}: recall@$K$ and ndcg@$K$, where $K$ is set as $20$ by default.
We report the average metrics for all users in the testing set.

\subsubsection{\textbf{Alternative Baselines}}
We compare KGIN with the state-of-the-art methods, covering KG-free (MF), embedding-based (CKE), and GNN-based (KGAT, KGNN-LS, CKAN, and RGCN) methods:
\begin{itemize}[leftmargin=*]
    \item \textbf{MF}~\cite{BPR} (matrix factorization) only considers the user-item interactions, while leaving KG untouched. Technically, it uses ID embeddings of users and items to perform the prediction.
    \item \textbf{CKE}~\cite{CKE} is a representative embedding-based method, which leverages KG embeddings of entities derived from TransR~\cite{TransR} as ID embeddings of items under the MF framework. Where, KG relations are only used as the constraints in TransR to regularize the representations of endpoints.
    \item \textbf{KGNN-LS}~\cite{KGNN-LS} is a GNN-based model, which converts KG into user-specific graphs, and then considers user preference on KG relations and label smoothness in the information aggregation phase, so as to generate user-specific item representations. It models relations in decay factors.
    \item \textbf{KGAT}~\cite{KGAT} is a state-of-the-art GNN-based recommender. It applies an attentive neighborhood aggregation mechanism on a holistic graph, which combines KG with the user-item graph, to generate user and item representations. User-item relationships and KG relations serve as the attentive weights in adjacent matrix.
    \item \textbf{CKAN}~\cite{CKAN} is built upon KGNN-LS, which utilizes different neighborhood aggregation schemes on the user-item graph and KG respectively, to obtain user and item embeddings.
    \item \textbf{R-GCN}~\cite{R-GCN} is originally proposed for the knowledge graph completion task, which views various KG relations as different channels of information flow when aggregating neighboring nodes. Here we transfer it to the recommendation task.
\end{itemize}

\subsubsection{\textbf{Parameter Settings}}
We implement our KGIN model in PyTorch, and have released our implementations (code, datasets, parameter settings, and training logs) to facilitate reproducibility.
For a fair comparison, we fix the size of ID embeddings $d$ as $64$, the optimizer as Adam~\cite{Adam}, and the batch size as $1024$ for all methods.
A grid search is conducted to confirm the optimal settings for each method --- more specifically, the learning rate is tuned in $\{10^{-4},10^{-3},10^{-2}\}$, the coefficients of additional constraints (\eg $L_{2}$ regularization in all methods, independence modeling in KGIN, TransR in CKE and KGAT, label smoothness in KGNN-LS) are searched in $\{10^{-5},10^{-4},\cdots,10^{-1}\}$, and the number of GNN layers $L$ is tuned in $\{1,2,3\}$ for GNN-based methods.
Moreover, for KGNN-LS and CKAN, we set the size of neighborhood as $16$ and the batch size as $128$.
We initialize model parameters with Xavier~\cite{Xavier}, while using the pre-trained ID embeddings of MF as the initialization of KGAT.

The detailed settings of KGIN are provided in Appendix~\ref{sec:reproducibility}.
We observe that using Equations \eqref{equ:independence-loss-mi} and \eqref{equ:independence-loss-dc} have similar trends and performance, hence report the results of Equation \eqref{equ:independence-loss-mi}.
We use KGIN-3 to denote the recommender model with three relational path aggregation layers, and similar notations for others.
Without specification, we fix the number of user intents $|\Set{P}|$ as $4$ and the number of relational path aggregation layers $L$ as $3$.
Moreover, in Sections~\ref{sec:user-intents} and~\ref{sec:model-depth}, we study their influence by varying $K$ in $\{1,2,4,8\}$ and $L$ in $\{1,2,3\}$, respectively.

\subsection{Performance Comparison (RQ1)}

\begin{table}[t]
    \caption{Overall performance comparison.}
    \centering
    \vspace{-10px}
    \label{tab:overall-performance}
    \resizebox{0.49\textwidth}{!}{
    \begin{tabular}{c|c c |c c| c c}
    \hline
    \multicolumn{1}{c|}{\multirow{2}*{}}& 
    \multicolumn{2}{c|}{Amazon-Book} & 
    \multicolumn{2}{c|}{Last-FM} &
    \multicolumn{2}{c}{Alibaba-iFashion} \\
      &recall & ndcg & recall & ndcg & recall & ndcg\\
    \hline
    \hline
    MF & 0.1300 & 0.0678& 0.0724& 0.0617& 0.1095 & 0.0670\\
    CKE & 0.1342 & 0.0698& 0.0732& 0.0630& \underline{0.1103} & \underline{0.0676}\\\hline
    KGAT & \underline{0.1487} & \underline{0.0799}& 0.0873& \underline{0.0744} & 0.1030 & 0.0627\\
    KGNN-LS & 0.1362 & 0.0560 & \underline{0.0880} & 0.0642 & 0.1039 & 0.0557\\
    CKAN & 0.1442 & 0.0698 & 0.0812 & 0.0660 & 0.0970 & 0.0509\\
    R-GCN & 0.1220 & 0.0646& 0.0743& 0.0631& 0.0860 & 0.0515\\
    \hline
    KGIN-3 & \textbf{0.1687$^*$} & \textbf{0.0915$^*$}& \textbf{0.0978$^*$}& \textbf{0.0848$^*$}& \textbf{0.1147$^*$} & \textbf{0.0716$^*$}\\
    \hline
    \hline
    \%Imp. & 13.44\% & 14.51\% & 11.13\% & 13.97\% & 3.98\% & 5.91\% \\
    \hline
    \end{tabular}}
    \vspace{-10px}
\end{table}

We begin with the comparison \wrt recall@$20$ and ndcg@$20$.
The empirical results are reported in Table~\ref{tab:overall-performance}, where \%Imp. denotes the relative improvements of the best performing method (starred) over the strongest baselines (underlined).
We find that:
\begin{itemize}[leftmargin=*]
    \item KGIN consistently outperforms all baselines across three datasets in terms of all measures. More specifically, it achieves significant improvements over the strongest baselines \wrt ndcg@$20$ by $14.51\%$, $13.97\%$, and $5.91\%$ in Amazon-Book, Last-FM, and Alibaba-iFashion, respectively.
    This demonstrates the rationality and effectiveness of KGIN.
    We attribute these improvements to the relational modeling of KGIN:
    (1) By uncovering user intents, KGIN is able to better characterize the relationships between users and items, and result in more powerful representations of users and items.
    In contrast, all baselines ignore the hidden user intents, and model user-item edges as a homogeneous channel to collect information;
    (2) Benefiting from our relational path aggregation scheme, KGIN can preserve the holistic semantics of paths and collect more informative signals from KG, than the GNN-based baselines (\ie KGAT, CKAN, KGNN-LS);
    (3) Applying different aggregation schemes on IG and KG makes KGIN better able to encode the collaborative signals and item knowledge into user and item representations.

    \item Jointly analyzing KGIN across the three datasets, we find that the improvement on Amazon-Book is more significant than that on Alibaba-iFashion. This is reasonable since (1) both interaction and KG data on Amazon-Book offer denser and richer information than that on Alibaba-iFashion; and (2) in Alibaba-iFashion, the first-order connectivity (\emph{fashion outfit, including, fashion staff}) dominate the KG triplets. This indicates that KGIN is good at fulfilling the potentials of long-range connectivity.
    % Another possible reason is that the relation distribution in Alibaba-iFashion is extremely skewed, which is dominated by the relation \emph{belong-to}. This indicates that KGIN possibly fulfils the potential of rich relational data.
    
    \item Leaving KG untapped limits the performance of MF. By simply incorporating KG embeddings into MF, CKE performs better than MF. Such findings are consist to prior studies~\cite{KTUP}, indicating the importance of side information like KG.
    
    \item GNN-based methods (\ie KGAT, CKAN, KGNN-LS) outperform CKE in Amazon-Book and Last-FM, suggesting the importance of modeling long-range connectivity. These improvements come from using the local structure of a node --- more specifically, multi-hop neighborhood --- to improve the representation learning. However, in Alibaba-iFashion, the performance of CKE is better than them. Some possible reasons are: (1) These GNN-based methods involve additional nonlinear feature transformation, which are rather heavy and burdensome to train, thus degrades performance~\cite{SGCN,LightGCN}; (2) TransR in CKE successfully captures the major first-order connectivity in Alibaba-iFashion.

    \item The results of KGAT, KGNN-LS, and CKAN are at the same level, being better than R-GCN.
    Although the utility of transforming neighbors' information via KG relations in R-GCN is better than working as decay factors in the others, R-GCN is not originally designed for recommendation and thus fails to model user-item relationships properly.
    
\end{itemize}

\subsection{Relational Modeling of KGIN (RQ2)}

\begin{table}[t]
    \caption{Impact of presence of user intents and KG relations.}
    \centering
    \vspace{-10px}
    \label{tab:impact-of-relations-presence}
    \resizebox{0.465\textwidth}{!}{
    \begin{tabular}{l|c c |c c| c c}
    \hline
    \multicolumn{1}{c|}{\multirow{2}*{}}& 
    \multicolumn{2}{c|}{Amazon-Book} & 
    \multicolumn{2}{c|}{Last-FM} &
    \multicolumn{2}{c}{Alibaba-iFashion} \\
      &recall & ndcg & recall & ndcg & recall & ndcg\\
    \hline
    \hline
    w/o I\&R  & 0.1518& 0.0816& 0.0802& 0.0669& 0.0862& 0.0530\\
    w/o I & 0.1627& 0.0870& 0.0942& 0.0819& 0.1103& 0.0678\\
    \hline
    \end{tabular}}
    \vspace{-5px}
\end{table}

\begin{table}[t]
    \caption{Impact of the number of layers $L$.}
    \centering
    \vspace{-10px}
    \label{tab:impact-of-model-depth}
    \resizebox{0.465\textwidth}{!}{
    \begin{tabular}{c|c c |c c| c c}
    \hline
    \multicolumn{1}{c|}{\multirow{2}*{}}& 
    \multicolumn{2}{c|}{Amazon-Book} & 
    \multicolumn{2}{c|}{Last-FM} &
    \multicolumn{2}{c}{Alibaba-iFashion} \\
      &recall & ndcg & recall & ndcg & recall & ndcg\\
    \hline
    \hline
    KGIN-1 & 0.1455 & 0.0766& 0.0831& 0.0707& 0.1045 & 0.0638\\
    KGIN-2 & 0.1652 & 0.0892& 0.0920& 0.0791& 0.1162 & 0.0723\\
    KGIN-3 & 0.1687 & 0.0915& 0.0978& 0.0848& 0.1147 & 0.0716\\
    \hline
    \end{tabular}}
    \vspace{-10px}
\end{table}

As the relational modeling is at the core of KGIN, we also conduct ablation studies to investigate the effectiveness --- specifically, how the presence of user intents and KG relations, the number of relational path aggregation layers, the granularity of user intents, and the independence of user intents influence our model.

\subsubsection{\textbf{Impact of Presence of User Intents \& KG Relations}}\label{sec:user-intents}
We first answer the question: Is it of importance to consider user intents or KG relations?
Towards this end, two variants are constructed by (1) discarding all user intents and KG relations, termed KGIN-3$_{\text{w/o I\&R}}$, and (2) removing all user intents only ($|\Set{P}|=0$), termed KGIN-3$_{\text{w/o I}}$.
We summarize the results in Table~\ref{tab:impact-of-relations-presence}.

Obviously, compared with KGIN-3 in Table~\ref{tab:overall-performance}, removing all relations (\ie KGIN-3$_{\text{w/o I\&R}}$) dramatically reduces the predictive accuracy, indicating the necessity of relational modeling.
To be more specific, KGIN-3$_{\text{w/o I\&R}}$ only propagates nodes' information in one space, without preserving any relational semantics, thus distorts the internal relationships among nodes.
Analogously, leaving hidden user intents unexplored (\ie KGIN-3$_{\text{w/o I}}$) also downgrades the performance.
Although KGIN-3$_{\text{w/o I}}$ retains the modeling of KG relations, it only considers coarser-gained preference of users and thus leads to suboptimal user representations.

\begin{table}[t]
    \caption{Impact of independence modeling.}
    \centering
    \vspace{-10px}
    \label{tab:impact-of-independence}
    \resizebox{0.48\textwidth}{!}{
    \begin{tabular}{c|cc|cc|cc}
    \hline
     & \multicolumn{2}{c|}{Amazon-Book} & \multicolumn{2}{c|}{Last-FM} & \multicolumn{2}{c}{Alibaba-iFashion} \\ 
     & w/ Ind & w/o Ind & w/ Ind & w/o Ind & w/ Ind & w/o Ind \\ \hline\hline
    distance correlation & 0.0389 & 0.3490 & 0.0365 & 0.4944 & 0.0112 & 0.3121 \\ \hline
    \end{tabular}}
    \vspace{-5pt}
\end{table}

\begin{figure}[t]
	\centering
	\subcaptionbox{Amazon-Book\label{fig:gamma-hit-mi}}{
		\includegraphics[width=0.48\columnwidth]{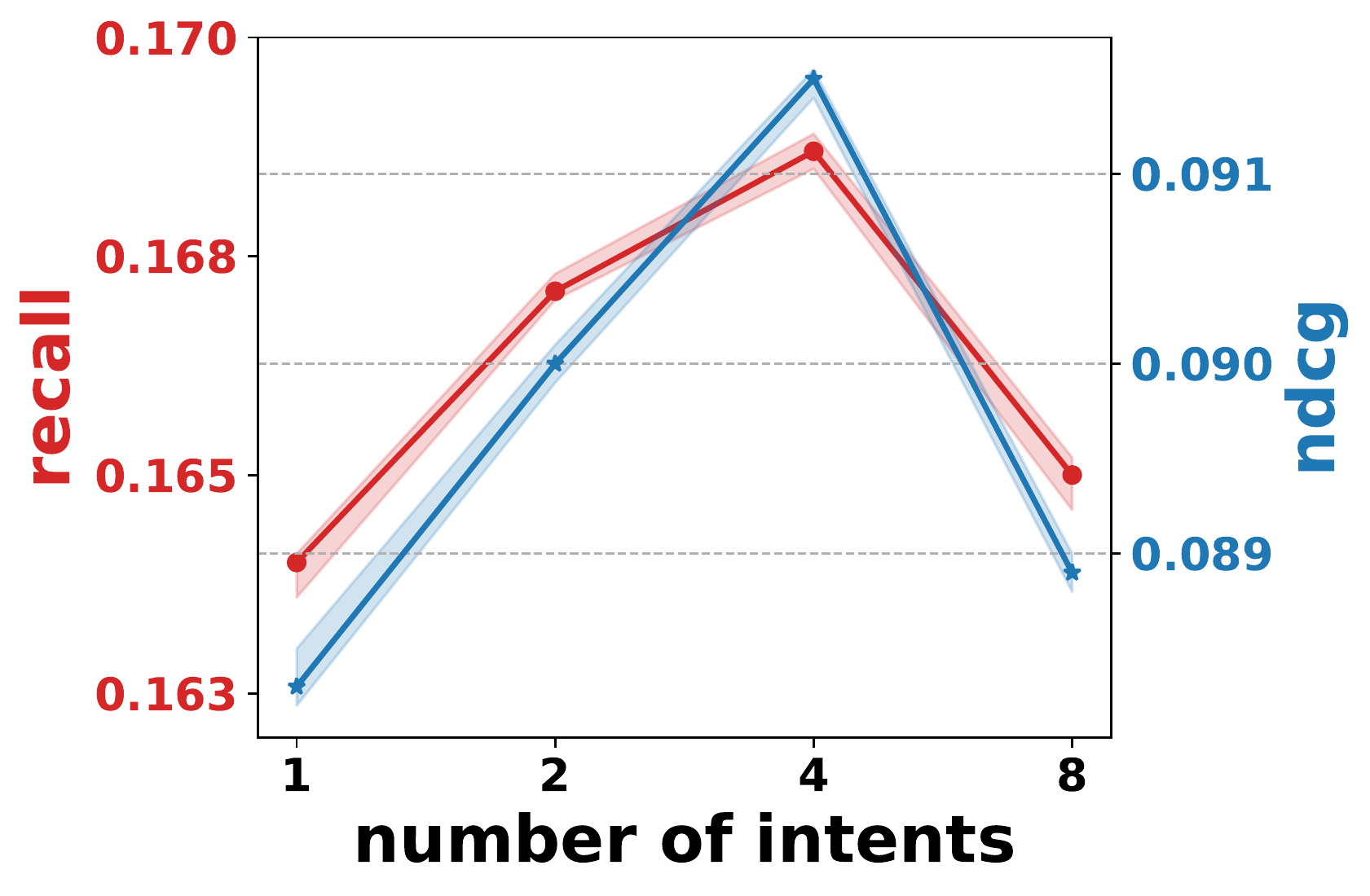}}
	\subcaptionbox{Last-FM\label{fig:gamma-ndcg-mi}}{
		\includegraphics[width=0.48\columnwidth]{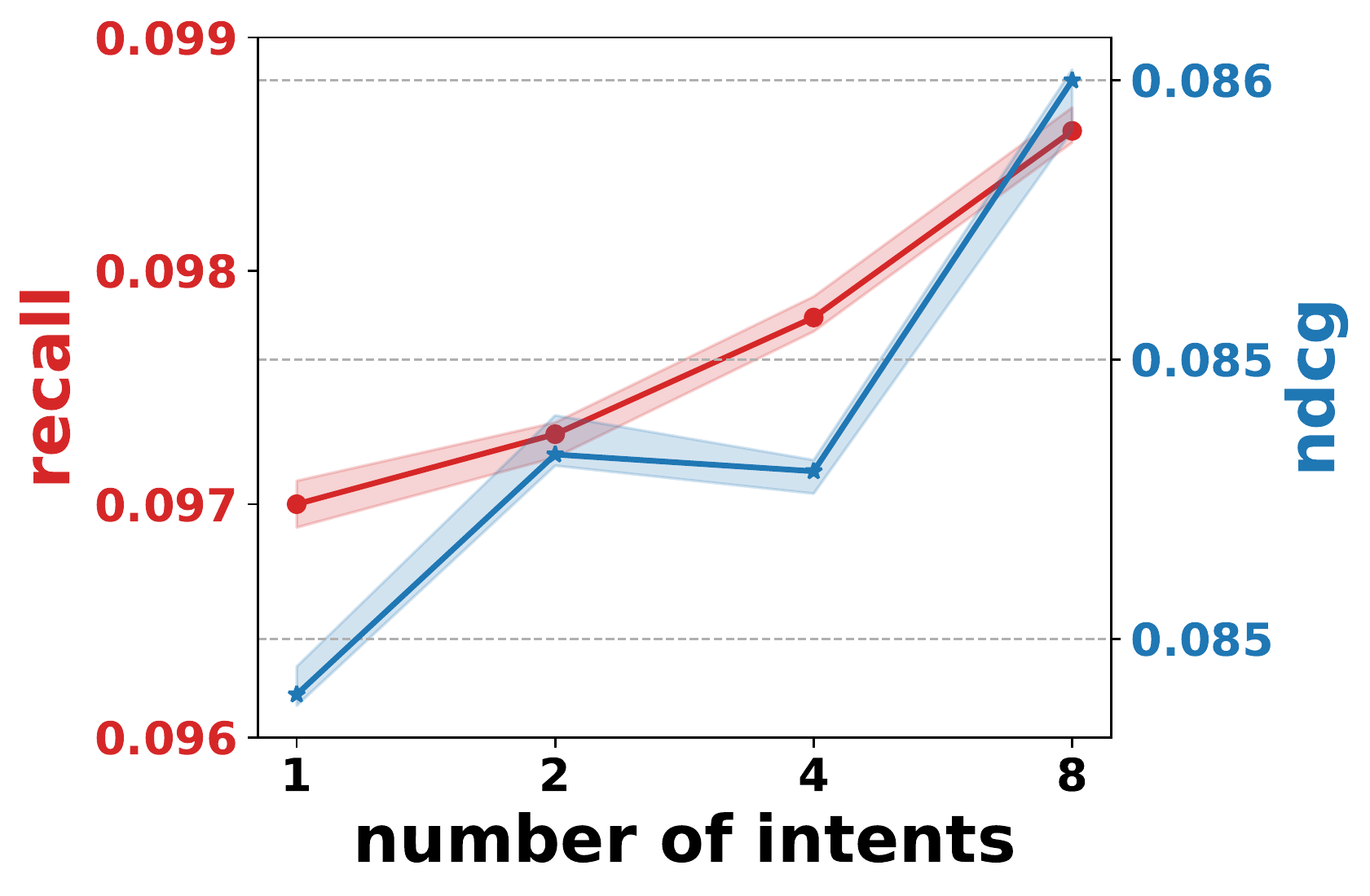}}
	\vspace{-10pt}
	\caption{Impact of intent number ($|\Set{P}|$). Best viewed in color.}
	\label{fig:impact-of-intent-number}
	\vspace{-10pt}
\end{figure}

\subsubsection{\textbf{Impact of Model Depth}}\label{sec:model-depth}
We then consider varying the number of relational path aggregation layers.
Stacking more layers is able to integrate the information carried by longer-range connectivity (\ie longer paths) into node representations.
Here we search $L$ in the range of $\{1,2,3\}$ and summarize the results in Table~\ref{tab:impact-of-model-depth}. 
We observe that:
\begin{itemize}[leftmargin=*]
    \item Increasing the model depth is able to enhance the predictive results in most cases. To be more specific, KGIN-2 substantially achieves significant improvements over KGIN-1. We attribute such improvements to two reasons: (1) Stacking more layers explore more relevant items connected by some KG triplets and deepens the understanding of user interest. KGIN-1 only takes the first-order connectivity (\eg user-intent-item triplets, KG triplets) into consideration, while KGIN-2 reveals the two-hop paths;
    (2) More information pertinent to user intents are derived from longer relational paths, thus better profiling user preference on items.

    \item Continuing one more exploration beyond KGIN-2, the results of KGIN-3 are consistently better in Amazon-Book and Last-FM. This empirically shows that higher-order connectivity is complementary to the second-order one, thus resulting in better node representations.
    
    \item However, the results of KGIN-3 are worse than KGIN-2 in Alibaba-iFashion. This again admits the inherent characteristics of Alibaba-iFashion --- most of KG triplets are the first-order connectivity (\emph{fashion outfit, including, fashion staff}) of items, which have been captured in KGIN-2.
    
    % \item Jointly analyzing the results in Tables~\ref{tab:overall-performance} and \ref{tab:impact-of-model-depth}, KGIN-2 and KGIN-3 are superior to the other baselines consistently. This again verifies the effectiveness of relational path aggregation mechanism. 

\end{itemize}

\begin{figure*}[ht]
    \centering
	\includegraphics[width=0.95\textwidth]{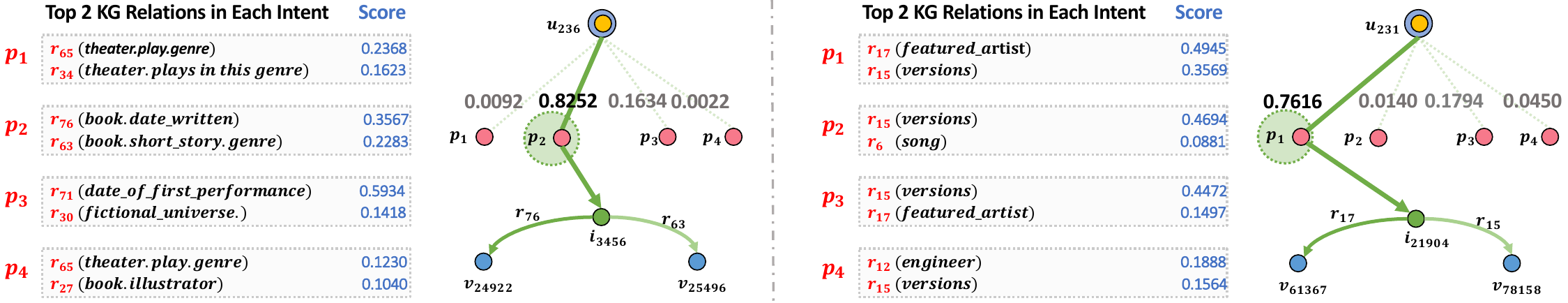}
	\vspace{-10pt}
	\caption{Explanations of user intents and real cases in Amazon-Book (left) and Last-FM (right). Best viewed in color.}
	\label{fig:case-study}
	\vspace{-10pt}
\end{figure*}

\subsubsection{\textbf{Impact of Intent Modeling}}
To analyze the influence of intents number, we vary $|\Set{P}|$ in range of $\{1,2,4,8\}$ and illustrate the performance changing curves on Amazon-Book and Last-FM datasets in Figure~\ref{fig:impact-of-intent-number}. We find that:
\begin{itemize}[leftmargin=*]
    \item Increasing the intent number enhances the performance in most cases. Specifically, when only modeling a coarse-grained relation (\ie $|\Set{P}|=1$), KGIN-3 performs poor across the board. This again emphasizes the benefits of exploring multiple user intents.
    \item In Amazon-Book, continuing one more partition beyond $|\Set{P}|=4$ impairs the accuracy. One possible reason is that independence modeling encourages the irrelevance among intents, but also makes some intents too fine-grained to carry useful information. We leave the exploration of intent granularity to future work.
    \item Interestingly, comparing to the results on Amazon-Book, setting $|\Set{P}|=8$ improves the accuracy on Last-FM, although Amazon-Book contains richer set of KG relations as compared to Last-FM. We attribute this to the difference of two datasets. In particular, KG in Last-FM is converted from the attributes of albums, songs, and artists, while KG in Amazon-Book is extracted from Freebase and contains noisy relations irrelevant to user behaviors.
    
\end{itemize}

We also conduct an ablation study to investigate the influence of independence modeling (\cf Section~\ref{sec:independence-of-intents}).
Specifically, we disable this module to build a variant KGIN-3$_{\text{w/o Ind}}$, and show the results \wrt distance correlation in Table~\ref{tab:impact-of-independence}.
Clearly, while approaching the comparable performance of recommendation to KGIN-3, KGIN-3$_{\text{w/o Ind}}$ achieves larger correlation coefficients and fails to differentiate user intents, which are still opaque to understand user behaviors.

\subsection{Explainability of KGIN (RQ3)}
In this section, we present the semantics of user intents, and offer two examples of Amazon-Book and Last-FM to give an intuitive impression of our explainability.
As shown in Figure~\ref{fig:case-study}, we have the following observations:
\begin{itemize}[leftmargin=*]
    \item KGIN first induces intents --- the commonality of all users --- with various combinations of KG relations. For an intent, the weight of a relation reflects its importance to influence user behaviors. For example, in Amazon-Book, the top two relations of the first intent $p_{1}$ are \emph{theater.play.genre} and \emph{theater.plays.in-this-genre}, while \emph{date-of-the-first-performance} and \emph{fictional-universe} are assigned with the highest scores for the second intent $p_{3}$.
    Clearly, the learned intents abstract the shared reasons of user's choices.
    Moreover, thanks to the independence modeling, the intents tend to have distinct boundary, thus describing user behaviors from different and independent angles.
    However, $p_{1}$ and $p_{3}$ are highly relevant. This makes sense since only $9$ relations exist in Last-FM.

    \item It can be found that some relations get high weights in multiple intents, like \emph{version} in Last-FM. This indicates that such relations are common factors pertinent to user behaviors. Combining it with other relations like \emph{featured-artist}, KGIN induces the intent $p_{1}$ as the special version of music created by a certain artist.

    \item KGIN creates instance-wise explanations for each interaction --- the personalization of a single user. For the interaction $u_{231}$-$i_{21904}$ in Amazon-Book, KGIN searches the most influential intent $p_{1}$ based on the attention scores (\cf Equation~\eqref{equ:intent-attention}).
    Thus, it explains this behavior as \emph{User $u_{231}$ selects music $i_{21904}$ since it matches her interest on the featured artist and certain version}.
\end{itemize}

\section{Related Work}
Existing recommender models incorporated with KG roughly fall into four groups.

\vspace{5pt}
\noindent\textbf{Embedding-based Methods}~\cite{CKE,CFKG,KTUP,Chorus,DKN,DBLP:conf/sigir/HuangZDWC18} focus mainly on the first-order connectivity (\ie user-item pairs in interaction data, triplets in KG), hire KG embedding techniques (\eg TransE~\cite{TransE} and TransH~\cite{TransH}) to learn entity embeddings, and then use them as prior or context information of items to guide the recommender model.
For example, CKE~\cite{CKE} applies TransE on KG triplets, and feed the knowledge-aware embeddings of items into matrix factorization (MF)~\cite{BPR}.
KTUP~\cite{KTUP} employs TransH on user-item interactions and KG triplets simultaneously, to jointly learn user preference and perform KG completion.
% More recently, Chorus~\cite{Chorus} takes item relations with temporal dynamics to design relational representations of items, to improve the performance of sequential recommendation.
Although these methods demonstrate the benefits of knowledge-aware embeddings, they ignore the higher-order connectivity.
This make them fail to capture the long-range semantics or sequential dependencies of paths between two nodes, and thus limits their ability to uncover the underlying user-item relationships.

\vspace{5pt}
\noindent\textbf{Path-based Methods}~\cite{KPRN,DBLP:conf/recsys/Sun00BHX18,hu2018leveraging,DBLP:conf/recsys/CatherineC16,RippleNet,DBLP:conf/www/MaZCJWLMR19} account for the long-range connectivity by extracting paths that connect the target user and item nodes via KG entities.
Then these paths are used to predict user preference, such as via recurrent neural networks~\cite{KPRN,DBLP:conf/recsys/Sun00BHX18} and memory network~\cite{RippleNet}.
% For example, KPRN~\cite{KPRN} uses random walks to extract qualified paths and models user preference as path representations with a recurrent neural network.
For example, RippleNet~\cite{RippleNet} memorizes the item representations along with paths rooted at each user, and uses them to enhance user representations.
Clearly, the recommendation accuracy heavily relies on the quality of paths.
However, two mainstream path extraction methods suffer from some inherent limitations:
(1) Applying brute-force search easily leads to labor-intensive and time-consuming feature engineering, when large-scale graphs are involved~\cite{KPRN};
(2) When using meta-path patterns to filter path instances, it requires domain experts to predefine the domain-specific patterns, thus resulting in poor transferability to different domains~\cite{hu2018leveraging,NIRec}.

\vspace{5pt}
\noindent\textbf{Policy-based Methods}~\cite{PGPR,DBLP:conf/sigir/ZhouDC0RTH020,DBLP:conf/sigir/ZhaoWZZLX020,KERL,KGPolicy} get inspiration from recent success of reinforcement learning (RL), and design RL agents to learn path-finding policy.
For example, PGPR~\cite{PGPR} exploits a policy network to explore items of interest for a target user.
% KERL~\cite{KERL} encodes relational signals into the RL agent to assist the sequential recommendation.
These RL-based policy networks can be viewed as efficient and cheap alternatives to the brute-force search, which serve as the backbone models of conversational recommender systems \cite{DBLP:conf/kdd/LeiZ0MWCC20,DBLP:journals/corr/abs-2101-09459}.
However, the sparse reward signals, huge action spaces, and policy gradient-based optimization make these networks hard to train and converge to a stable and satisfying solution~\cite{DBLP:conf/emnlp/XiongHW17,DBLP:conf/sigir/ZhaoWZZLX020}.

\vspace{5pt}
\noindent\textbf{GNN-based Methods}~\cite{KGAT,KGCN,KGNN-LS,CKAN,NIRec} are founded upon the information aggregation mechanism of graph neural networks (GNNs)~\cite{DBLP:conf/nips/HamiltonYL17,DBLP:conf/iclr/KipfW17,DBLP:conf/iclr/VelickovicCCRLB18,NGCF,LightGCN}.
Typically, it incorporates information from the one-hop nodes to update the representations of ego nodes; when recursively performing such propagations, information from multi-hop nodes can be encoded in the representations.
As such, these methods are able to model long-range connectivity.
For instance, KGAT~\cite{KGAT} combines user-item interactions and KG as a heterogeneous graph, and then applies the aggregation mechanism on it.
CKAN~\cite{CKAN} uses two different strategies to separately spread collaborative signals and knowledge-aware signals.
More recently, NIRec~\cite{NIRec} is proposed to combine path- and GNN-based models, which propagates interactive patterns between two nodes through meta-path-guided neighborhoods.

However, to the best of our knowledge, current GNN-based methods assume that only one relation exists among users and items, but leave hidden intents unexplored.
Moreover, most of them fail to preserve the relational dependency in paths.
Our work differs from them in these relational modeling --- we focus on exhibiting user-item relationships at the granularity of intents, and encoding relational paths into the representations, towards better performance and interpretability. 

\section{Conclusion and Future Work}
In this work, we focused on the relational modeling of knowledge-aware recommendation, especially in GNN-based methods.
We proposed a novel framework, KGIN, which approaches better relational modeling from two dimensions:
(1) uncovering user-item relationships at the granularity of intents, which are coupled with KG relations to exhibit the explainable semantics;
and (2) relational path-aware aggregation, which integrates relational information from multi-hop paths to refine the representations.
We further offered in-depth analysis of KGIN \wrt the effectiveness and explainability of recommendation.

% Exploiting side information, especially the structural knowledge like knowledge graph, is a promising solution to enhancing the accuracy and explainability of recommendation.
Current works usually frame the KG-based recommendation as a supervised task, where the supervision signal comes from historical interactions only.
Such supervisions are too sparse to offer high-quality representations.
In future work, we will explore self-supervised learning in recommendation, in order to generate auxiliary supervisions via self-supervised tasks and uncover the internal relationships among data instances.
Furthermore, we would like to introduce causal concepts, such as causal effect inference, counterfactual reasoning, and deconfounding, into knowledge-aware recommendation to discover and amplify biases \cite{DBLP:journals/corr/abs-2010-03240}.

\begin{acks}
    This research is part of NExT++ research, which is supported by the National Research Foundation, Singapore under its International Research Centres in Singapore Funding Initiative. It is also supported by the National Natural Science Foundation of China (U19A2079, 61972372).
    This work is also supported by Janet Jenq, Heather Thibodeau and Beatriz Reyero del Rio of eBay Customer Science department and Mitch Wyle and Petra Hofer of eBay eRupt organization.
\end{acks}

\bibliographystyle{ACM-Reference-Format}
\balance
\bibliography{ms}
\balance

%%
%% If your work has an appendix, this is the place to put it.
\appendix
\section{Appendix}
\subsection{Reproducibility}\label{sec:reproducibility}

We list the parameter settings of KGIN on three datasets in Table~\ref{tab:parameter-settings}, where the hyperparameters include the learning rate $\rho$, the embedding size $d$, the number of aggregation layers $L$, the number of user intents $|\Set{P}|$, the coefficient $\lambda_{1}$ of the independence modeling $\Lapl_{\text{IND}}$, and the coefficient $\lambda_{2}$ of $L_{2}$ regularization.
We have released our codes, datasets, model parameters, and training logs at \url{https://github.com/huangtinglin/Knowledge_Graph_based_Intent_Network} to facilitate reproducibility.

\begin{table}[b]
    \caption{Hyperparameter settings of KGIN.}
    \vspace{-10px}
    \label{tab:parameter-settings}
    \resizebox{0.45\textwidth}{!}{
    \begin{tabular}{l|cccccc}
    \hline
     & $\rho$ & $d$ & $L$ & $|\Set{P}|$ & $\lambda_{1}$ & $\lambda_{2}$ \\ \hline\hline
    Amazon-Book & $10^{-4}$ & 64 & 3 & 4 & $10^{-5}$ & $10^{-5}$ \\ 
    Last-FM & $10^{-4}$ & 64 & 3 & 4 & $10^{-4}$ & $10^{-5}$ \\ 
    Alibaba-iFashion & $10^{-4}$ & 64 & 3 & 4 & $10^{-4}$ & $10^{-5}$ \\ \hline
    \end{tabular}}
\end{table}

\end{document}